\documentclass[useAMS,natbib,multicolumn,subfigure]{mn2e}
\usepackage{amsmath}
\usepackage[T1]{fontenc}
\usepackage{ae,aecompl}

\usepackage[latin1]{inputenc}  
\usepackage{graphicx,amsmath,xcolor}
\usepackage{amssymb}
\usepackage{indentfirst}
\usepackage{hyperref}
\usepackage{graphicx,pst-all}
\usepackage{tikz}
\usepackage{layout}
\usepackage{fancyhdr}
\usepackage{scrextend}

\newcommand{\kms}{\hbox{\kern 0.20em km\kern 0.20em s$^{-1}$}}
\newcommand{\cmt}{\hbox{\kern 0.20em cm$^{-3}$}}
\newcommand{\cmd}{\hbox{\kern 0.20em cm$^{-2}$}}

\def\msol{\hbox{\kern 0.20em $M_\odot$}}
\def\lsol{\hbox{\kern 0.20em $L_\odot$}}
\def\rsol{\hbox{\kern 0.20em $R_\odot$}}
\def\sr{\hbox{\kern 0.20em sr}}
\def\srmu{\hbox{\kern 0.20em sr$^{-1}$}}

\def\g{\hbox{\kern 0.20em g}}
\def\gmu{\hbox{\kern 0.20em g$^{-1}$}}
\def\kg{\hbox{\kern 0.20em kg}}
\def\pc{\hbox{\kern 0.20em pc}}

\def\mum{\hbox{\kern 0.20em $\mu$m}}
\def\mumd{\hbox{\kern 0.20em $\mu$m$^{-2}$}}
\def\cm{\hbox{\kern 0.20em cm}}
\def\m{\hbox{\kern 0.20em m}}
\def\km{\hbox{\kern 0.20em km}}
\def\nm{\hbox{\kern 0.20em nm}}

\def\s{\hbox{\kern 0.20em s}}
\def\h{\hbox{\kern 0.20em h}}
\def\sec{\hbox{\kern 0.20em sec}}
\def\min{\hbox {\kern 0.20em min}}
\def\smu{\hbox{\kern 0.20em s$^{-1}$}}
\def\smd{\hbox{\kern 0.20em s$^{-2}$}}
\def\an{\hbox{\kern 0.20em an}}
\def\anmu{\hbox{\kern 0.20em an$^{-1}$}}
\def\deg{\hbox{\kern 0.20em $^{\rm o}$}}
\def\yr{\hbox{\kern 0.20em yr}}
\def\yrmu{\hbox{\kern 0.20em yr$^{-1}$}}
\def\Myr{\hbox{\kern 0.20em Myr}}
\def\Mymu{\hbox{\kern 0.20em Myr$^{-1}$}}
\def\K{\hbox{\kern 0.20em K}}
\def\pcmu{\hbox{\kern 0.20em pc$^{-1}$}}
\def\pcmd{\hbox{\kern 0.20em pc$^{-2}$}}
\def\pcmt{\hbox{\kern 0.20em pc$^{-3}$}}
\def\kms{\hbox{\kern 0.20em km\kern 0.20em s$^{-1}$}}
\def\kmpd{\hbox{\kern 0.20em km$^{2}$}}
\def\kpc{\hbox{\kern 0.20em kpc}}
\def\cms{\hbox{\kern 0.20em cm\kern 0.20em s$^{-1}$}}
\def\erg{\hbox{\kern 0.20em erg}}
\def\ergs{\hbox{\kern 0.20em erg}}
\def\cmpd{\hbox{\kern 0.20em cm$^2$}}
\def\cmmd{\hbox{\kern 0.20em cm$^{-2}$}}
\def\cmms{\hbox{\kern 0.20em cm$^{-6}$}}
\def\cmpt{\hbox{\kern 0.20em cm$^3$}}
\def\cmmt{\hbox{\kern 0.20em cm$^{-3}$}}
\def\mpd{\hbox{\kern 0.20em m$^2$}}
\def\mmd{\hbox{\kern 0.20em m$^{-2}$}}
\def\mpt{\hbox{\kern 0.20em m$^3$}}
\def\mmt{\hbox{\kern 0.20em m$^{-3}$}}
\def\mujy{\hbox{\kern 0.20em $\mu$Jy}}
\def\mjy{\hbox{\kern 0.20em mJy}}
\def\Mj{\hbox{\kern 0.20em MJy}}
\def\jy{\hbox{\kern 0.20em Jy}}
\def\ghz{\hbox{\kern 0.20em GHz}}
\def\srmd{\hbox{\kern 0.20em sr$^{-1}$}}

\def \mum{$\mu$m}
\def\G{\hbox{\kern 0.20em G}}

\def\htwo{\hbox{H${}_2$}}
\def\h13cop{\hbox{H$^{13}$CO$^{+}$}}

\def\h2o{\hbox{H$_2$O}}

\begin{document}

\title[]{Molecules in the Cep\,E-mm jet: evidence for shock-driven photochemistry ?}

\author[J. Ospina-Zamudio]{J.~Ospina-Zamudio$^{1}$\thanks{E-mail: juan-david.ospina-zamudio@univ-grenoble-alpes.fr},
          B.~Lefloch$^{1}$,
          C.~Favre$^{1}$,
          A.~L\'opez-Sepulcre$^{1,2}$,
          E.~Bianchi$^{1}$,
          \newauthor
          C.~Ceccarelli$^{1}$,
          M.~DeSimone$^{1}$,
          M.~Bouvier$^{1}$,
          C.~Kahane$^{1}$
\\
$^1$CNRS, IPAG, Univ. Grenoble Alpes, F-38000 Grenoble, France
\\
$^2$Institut de Radioastronomie Millim\'{e}trique (IRAM), 38406 Saint Martin d'H\`{e}res, France}
\date{Accepted 2019 September 25. Received 2019 September 25; in original form 2019 August 19}
\pubyear{2019}

\label{firstpage}
\pagerange{\pageref{firstpage}--\pageref{lastpage}}
\maketitle

\begin{abstract}
The chemical composition of protostellar jets and its origin  are still badly understood. More observational constraints are needed to make progress. With that objective, we have  carried  out a systematic search for molecular species in the  jet of  Cep\,E-mm, a template for intermediate-mass Class 0 protostars, associated with a luminous, high-velocity outflow.  We made use of  an unbiased spectral line survey in the range 72-350 GHz obtained with the IRAM 30m telescope, complementary observations of the CO $J$=3--2 transition with the JCMT, and  observations at $1\arcsec$ angular resolution of the   CO $J$=2--1 transition  with the IRAM Plateau de Bure interferometer. In addition to CO, we have detected rotational transitions from  SiO, SO, H$_2$CO, CS, HCO$^{+}$ and HCN. A strong chemical differentiation is observed in the  southern and northern lobes of the jet. Radiative transfer analysis in the Large Velocity Gradient approximation yields typical molecular abundances of the order of $10^{-8}$ for all molecular species other than CO. Overall, the jets exhibit an unusual chemical composition, as CS, SO and H$_2$CO are found to be the most abundant species, with a typical abundance of (3--4)$\times 10^{-8}$. The transverse size of the CO jet emission estimated from interferometric observations is about 1000 au, suggesting that  we are detecting emission from a turbulent layer of gas entrained by the jet in its propagation and not the jet itself. We propose that some  molecular species could be  the signatures of the specific photochemistry driven by the UV radiation field generated in the turbulent envelope. 
\end{abstract}


\maketitle
\section{Introduction}
Protostellar outflows are one of the most spectacular manifestations of the mass-loss phenomena
which take place along the protostellar evolution, from the early, embedded  Class 0 to the late
Class I phase. Many protostellar outflows show evidence for molecular jets.  The origin of molecules in protostellar jets is strongly debated and observational constraints are lacking. This issue is all the more important as the jet  molecular composition  could bring useful constraints on  the gas dynamics  and possibly on the launching mechanism itself. Numerical simulations suggest that turbulent entrainment from the surrounding envelope is not efficient enough to explain the CO column densities (Taylor et al. 1995). Several other mechanisms have been proposed: formation by ion chemistry at the base of an atomic stellar wind (Glassgold et al. 1991), formation behind atomic jet shocks (Raga et al. 2005), direct MHD ejection of molecules from the disk (Panoglou et al. 2012).  Therefore, a census of the molecular composition of protostellar jets could help discriminate between these various scenarii. 

The Large Program CALYPSO has investigated the emission of a few molecular tracers between 92 and 232 GHz with the IRAM Plateau de Bure interferometer towards a sample of 16 nearby ($d \leq 300\pc$) Class 0 protostars objects (Maury et al. 2015). It comes out that jets are often associated with SiO and SO, in addition to CO (see Codella et al. 2014; Santangelo et al. 2015). Until now, there have been very few attempts to investigate the molecular content of these jets.  Tafalla \& Hacar (2010) have led a molecular survey of the content of the two protostellar outflows L1448-mm and IRAS~04166+2706, from which they reported the detection of SiO, SO, HCN, and CH$_3$OH. The molecular composition differs markedly between the low-velocity bipolar outflow and the protostellar jet itself. This point is well illustrated by the L1157 low-velocity outflow which was long identified as "chemically active" (Bachiller \& Perez-Gutierrez 1997; Codella et al. 2010; Lefloch et al. 2017) whereas the jet molecular emission  was  detected only recently, in the lines of SiO and CO  (Tafalla et al. 2015; Podio et al. 2016).

In this context, we have conducted an unbiased molecular line survey of the high-velocity jet and outflow of Cep\,E-mm, a template for intermediate-mass Class 0 protostar. The source has a luminosity of about $100\lsol$ (Lefloch et al. 1996, Chini et al. 2001) and is located at a distance of 730~pc in the Cepheus-E molecular cloud (Sargent 1977).
Both the source and its outflow have been extensively observed in the millimetre to far-infrared wavelengths range (see, e.g., Eisl\"offel et al. 1996; Lefloch et al. 1996;  Noriega-Crespo et al. 1998; Hatchell et al. 1999b; Moro-Martin et al 2001;  Smith et al. 2003; Lefloch et al. 2015; Ospina-Zamudio et al. 2018).
However the molecular content of Cep\,E-mm have received only little attention until the systematic study by Ospina-Zamudio et al. (2018). These authors have reported the detection of a hot corino in the source. They have shown that Cep\,E-mm is actually a binary system, whose components  Cep\,E-A and Cep\,E-B both drive high-velocity molecular jets. The authors identified Cep\,E-A as the powering source of the high-velocity jet associated with HH377.
They also found  that the outflow is chemically active, following the empirical classification proposed by Bachiller et al. (1997). The physical properties of the southern lobe of the outflow, both the entrained gas and the jet, were obtained from a  multi-line analysis  between $J$=1--0 and $J$=16--15, of the CO jet emission (Lefloch et al. 2015). This study revealed the presence of hot, dense CO knots along the jet, which were interpreted as the signature of internal shocks.

In this article, we report the results of our unbiased search for molecular emission in  the high-velocity jet of Cep\,E-A. The article is organized as follows. In Section 2, we present the observations carried out with the IRAM~30m telescope, the JCMT and the IRAM Plateau de Bure Interferometer (PdBI).  In Section 3, we present the results on the jet molecular content. In addition to the previously known  molecules in the jet (CO, SiO), we report on the detection of SO, CS, H$_2$CO, HCO$^{+}$ and HCN. In Section 4, we derive the physical conditions and the molecular abundances of the various tracers detected in the jet. In each ouflow lobe, the jet composition is compared with that of the low-velocity outflow cavity. We discuss the origin of the jet molecular emission. The  specific jet molecular composition is found to agree qualitatively with the predictions of a simple  shock-driven UV photochemical model. Our conclusions are summarized in Section 5.

\section{Observations}
\subsection{IRAM 30m}

The observational data towards the protostar on coordinates $\alpha(2000)=\mathrm{23^h03^m12^s.8},\,\delta(2000)=\mathrm{61^{\circ}42'26''}$ have already been presented in Ospina-Zamudio et al. (2018). 
As recalled by the authors, calibration uncertainties in the 3, 2, 1.3 and 0.9mm bands are typically 10, 15, 20 and 30\%, respectively.
In this work, line intensities are expressed in units of antenna temperature corrected for the atmospheric absorption $T_{A}^*$. The rms noise per velocity interval of $1\kms$ expressed in units of  $T_{A}^*$ lies in the range 2--5~mK in the 3mm band, 3--7~mK in the 2mm and 1.3mm band, and 15--20~mK in the 0.9mm band.

The observational parameters and the parameters of all the detected lines are given in Table~1.
Molecular line brightness temperatures $T_{\rm B}$  were estimated from main-beam temperatures $T_{\rm MB}$, following the procedure described in Lefloch et al. (2015) to estimate the filling factor of the molecular jet emission in the telescope main-beam. We adopted the main-beam efficiency values $\eta_{\rm mb}$, as monitored by IRAM\footnote{http://www.iram.fr}.

\subsection{JCMT}
The CO lines $J$=3--2 345.796 GHz and $J$=4--3 461.041 GHz were observed toward the protostar position at the JCMT in June 1997. The CO $J$=4--3 was also observed at the offset position ($0\arcsec$,$+12\arcsec$). The observational procedures and data reduction procedures have been discussed in Lefloch et al. (2015).

\subsection{PdBI}

The CO $J$=2--1 line at 230.538~GHz was observed with the five-antenna array of the PdBI in March 1997 in configurations C and D, and again in January and February 1998 in configurations A and B.
The details of the observing procedure and data reduction have been presented in Lefloch et al. (2015). As a result of the calibration conducted under CLIC\footnote{http://www.iram.fr/IRAMFR/GILDAS/}, we obtained a natural weighted map with a synthetic beam of 1.07" $\times$ 0.87" (PA=73$^o$).

\section{The high-velocity jet}

\begin{figure}
\centering
\includegraphics[width=\columnwidth]{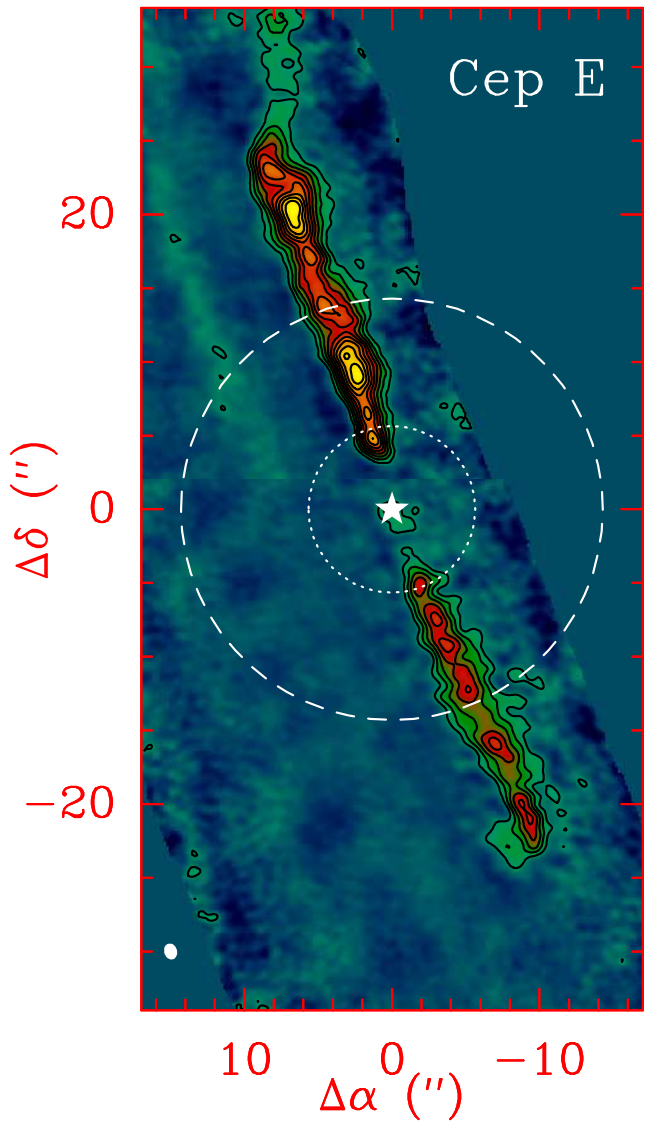}
\caption{CO $J$=2--1 velocity-integrated emission of the protostellar jet observed with the PdBI at $1\arcsec$ resolution (from Lefloch et al. 2015). The emission is integrated between $-135$ and $-110\kms$ and between $+50$ and $+80\kms$ towards the southern and northern lobes, respectively.  The first contour and contour interval are 20\% and 10\% of the peak intensity. The synthetic beam size (HPBW) is drawn by a filled ellipse. The IRAM 30m telescope beam size (HPBW) at the frequency of the SiO $J$=2--1 and $J$=5--4 is drawn by a dashed and dotted circle, respectively. A white star marks the position of the protostar Cep\,E-A. }
    \label{panel_line}
\end{figure}

\begin{figure}
\centering
\includegraphics[width=\columnwidth]{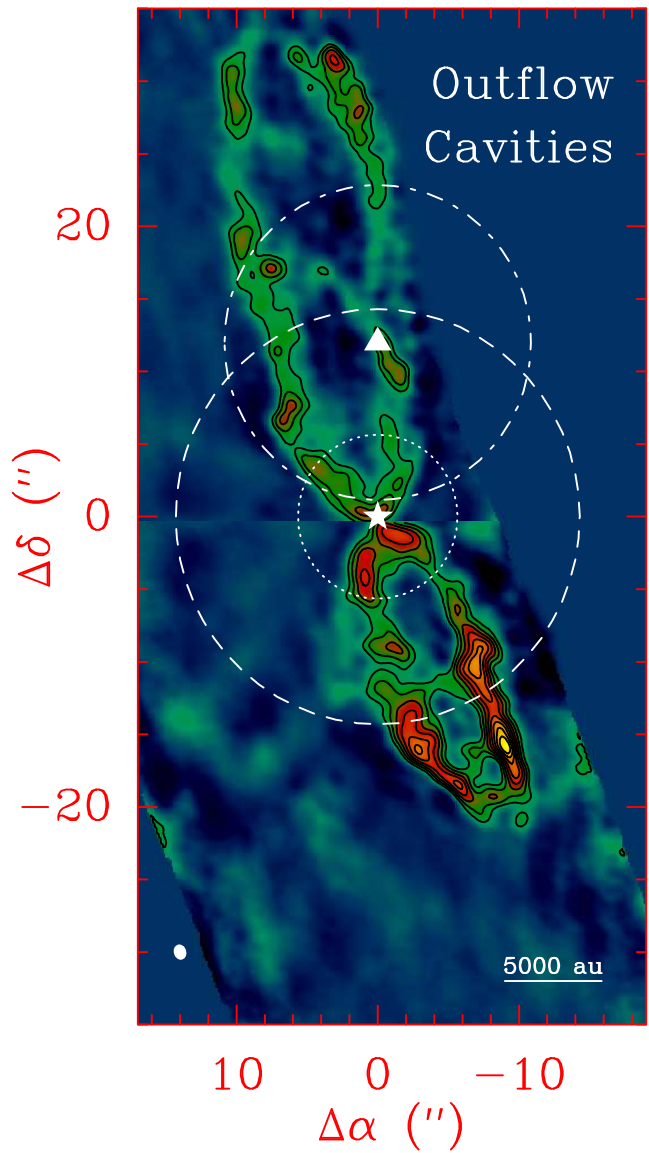}
\caption{CO $J$=2--1 velocity-integrated emission of the  low-velocity outflow observed with the PdBI at $1\arcsec$ resolution. The emission is integrated between $-20$ and $-14\kms$ and between $-8$ and $-6\kms$ towards the southern and northern lobes, respectively. The first contour and contour interval are 20\% and 10\% of the peak intensity. The synthetic beam size (HPBW) is drawn by a filled ellipse. The IRAM 30m telescope beam size (HPBW)  at the frequency of the SiO $J$=2--1 and $J$=5--4 is drawn by a dashed and dotted circle, respectively. The filled triangle marks the offset position ($0\arcsec$,$+12\arcsec$) observed with the IRAM 30m and the JCMT. The dashed-dotted circle draw the IRAM 30m beam at the frequency of the CO $J$=1--0 line. A white star marks the position of the protostar Cep\,E-A.}
    \label{panel_line}
\end{figure}

\begin{figure}
\centering
\includegraphics[width=\columnwidth]{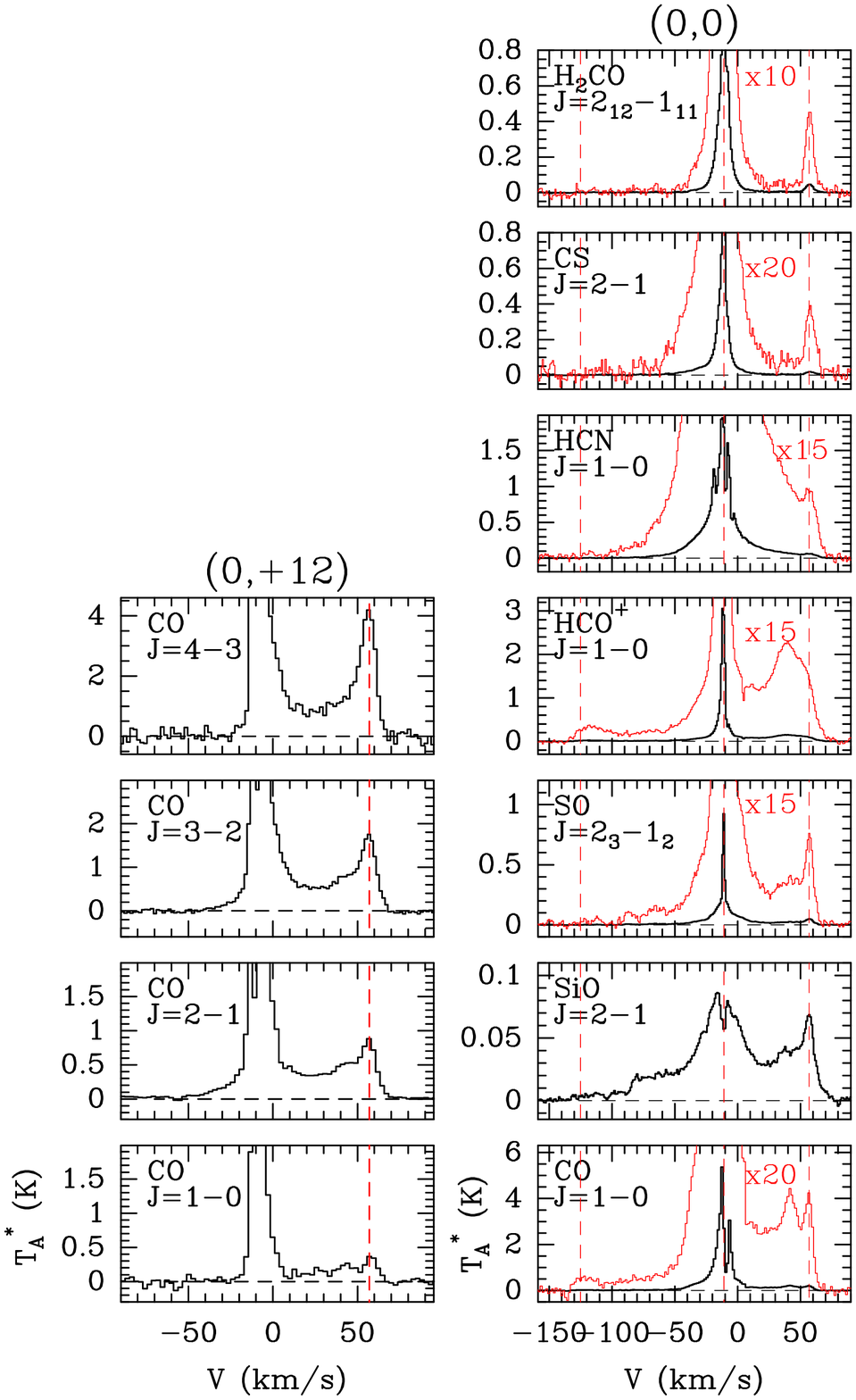}
\caption{({\em left})~CO line emission detected towards the outflow northern lobe at the offset position ($0\arcsec$,$+12\arcsec$). ({\em right})~Molecular line emission detected towards the protostarposition (0$\arcsec$,0$\arcsec$). Original spectra are displayed in black. Spectra with a magnifying factor are drawn in thin red lines.  Fluxes are expressed in units of antenna temperature $T_{\rm A}^{*}$, corrected for atmospheric absorption. The ambient cloud velocity ($v_{\rm lsr}$= $-10.9\kms$) and the mean jet velocity in the northern and southern lobes ($v_{\rm lsr}$= $+57\kms$ and $-125\kms$, respectively) are marked by red, dashed lines.
}
    \label{panel_line}
\end{figure}

Analysis of the CO $J$=2--1 interferometric observations  (Lefloch et al. 2015) revealed three spatial components in the protostellar outflow: the molecular jet (Fig.~1) , the outflow cavity (Fig.~2), and the terminal shock HH377.
Each of these components appear to have a specific spectral signature.\\
The molecular jet (hereafter the high-velocity jet) powered by the protostar Cep\,E-A (Ospina-Zamudio et al. 2018) is detected between $v_{\rm lsr}$= $-135$ and $-110\kms$ and between $+50$ and $+80\kms$ towards the southern and northern lobes, respectively (Fig.~1).\\
The outflow cavity is detected at velocities from ambient  ($v_{\rm lsr}$= $-10.9\kms$) up to $-100\kms$ ($+60\kms$) in the southern (northern) lobe. The bulk of the emission is detected up to $10\kms$ from the ambient cloud velocity (hereafter the low-velocity outflow) and arises from the outflow cavity walls (Fig.~2).

A systematic search down to the $3\sigma$ level yielded detection of the following molecular species in the high-velocity jet: CO, SiO, SO, HCO$^{+}$, HCN, CS and H$_2$CO (Fig.~3). All the molecular transitions detected in the jet are listed in Table~1 and a montage of the lowest excitation transitions of each molecular species is displayed in the right panel of Fig.~3. The spectra of all the transitions detected are displayed in Figs.~A1--A4. Inspection of the line profiles in Fig.~3 and Figs.~A1--A4 immediately show:
\begin{itemize}
\item different chemical properties between both jet lobes: For instance,  H$_2$CO and CS lines are bright in the northern lobe, while these species are not  detected in the southern lobe.  On the contrary, HCO$^+$ is detected in both lobes. 

\item different physical structures:   different molecular species display different line profiles inside the same lobe and between both lobes.
\end{itemize}

\subsection{CO}

The  single-dish CO line observations  towards the offset position $(0\arcsec$,$+12\arcsec$) and 
($0\arcsec$,$0\arcsec$ ) show that the CO emission ranges between $-150\kms$ and $+80\kms$ (Fig.~3).  The complex structure of the outflowing gas is evidenced  by the presence of several high-velocity components in both outflow lobes,
and the various spectral signatures detected with different molecular gas tracers.
Whereas the ambient cloud emission peaks at $v_{\rm lsr}$=$-10.9\kms$, the  emission of the protostellar jet
lies in the velocity intervals [$+50$;$+80$]$\kms$ and [$-150$;$-100$]$\kms$ for the redshifted and blueshifted lobe, respectively and peaks at $+56\kms$ and $-125\kms$, respectively(see also Lefloch et al. 1996, 2015). A map of the
CO $J$=2--1 jet velocity-integrated emission between [$+50$;$+80$]$\kms$ and [$-150$;$-100$]$\kms$, respectively, obtained with the PdBI at $1\arcsec$ resolution is displayed in Fig.~1. The overall emission of both jet lobes is rather similar similar, with a clumpy structure suggestive of internal shocks and multiple ejections (see also Lefloch et al. 2015). We note that the CO $J$=2--1 line emission brightness in the northern lobe is larger than in the southern lobe by typically  a factor of $\sim 5$.

\subsection{SiO}

\begin{figure}
\centering
\includegraphics[width=0.7\columnwidth]{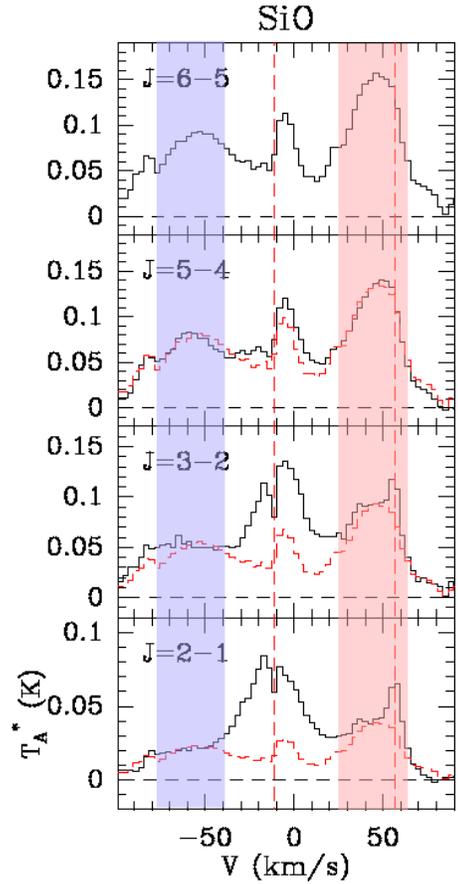}
\caption{Montage of the SiO transitions $J$=2--1, 3--2, 5--4 and 6--5, detected with the IRAM 30m telescope. The ambient cloud velocity ($v_{\rm lsr}$=$-10.9\kms$) and the jet emission peak velocity in the northern lobe ($+56\kms$) are marked with red dashed lines. The SiO $J$=6--5 line profile is superimposed in dashed red on the different SiO spectra. A scaling factor has been applied to the $J$=6--5 line profile so to match the emission of the high-velocity gas in the intervals marked by red and blue filled background rectangles. Fluxes are expressed in units of antenna temperature $T_{\rm A}^{*}$, corrected for atmospheric absorption.}

    \label{panel_line}
\end{figure}

Outflow emission was detected in all the SiO transitions present in the survey. A montage of the different transitions is displayed in Fig.~A1.  
The jet signature in the northern lobe is unambiguously detected at $v$= $+56\kms$  in the SiO $J$=2--1 and 3--2  transitions (Figs.~3 and 4). We failed to detect jet emission in the higher-$J$ ($J$ > 3) transitions. This is probably because the telescope beam size (HPBW) is less than $11\arcsec$ at the frequency of the high-excitation lines of SiO, hence, it encompasses a very small emitting region from the jet (see Fig.~1). Also, as can be seen in Fig.~4, the profiles of the transitions $J$=5--4 and 6--5  show an excellent agreement once they are scaled to match the intensity of the high-velocity range. More generally, for all SiO transitions, a very good match is observed  for the both blue- and redshifted gas in the velocity intervals [$-80$;$-40$]$\kms$ and [$+25$;$+62$]$\kms$, respectively (blue and red filled areas in Fig.~4). This suggests that the different transitions are tracing a component of high-excitation, which spans a broad velocity range. Since the line profile is independent of the transition and the telescope beam size, we conclude that the size of this component must be small in front of the telescope beam.

As illustrated in Fig.~4, the $J$=2--1 and 3--2 line profiles can be decomposed as the sum of the high-excitation, broad line component traced by the $J$=6--5 and 5--4 lines, and a "narrow" gaussian line  (FWHM$\simeq$ $9\kms$), which peaks at the  northern jet velocity ($+56\kms$). We identify the signature of the jet with this narrow line component.

\subsection{SO}
\begin{figure}
\centering
\includegraphics[width=\columnwidth]{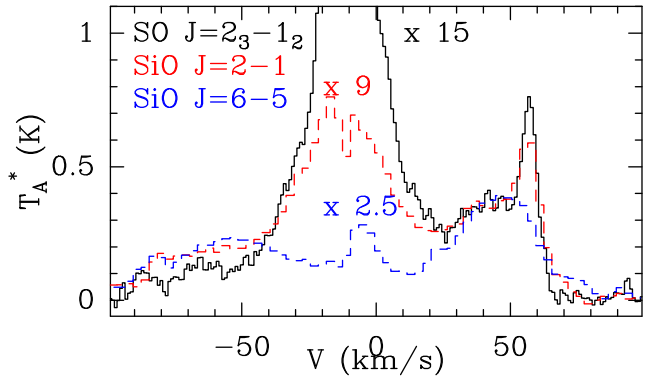}
\caption{Comparison of the profiles of SO $J$=$2_3$--$1_2$ transition (solid black) with SiO $J$=2--1 (dashed red) and SiO J=6--5 (dashed blue). Magnifying factors of 15, 9 and 2.5, respectively, have been applied in order to enhance and match the high-velocity emission in the jet. Fluxes are expressed in units of antenna temperature $T_{\rm A}^{*}$, corrected for atmospheric absorption.}
\end{figure}
For SO, we detected all the transitions with an Einstein coefficient $A_{\rm ij}$ larger than $10^{-5}\smu$ in the 3mm band and $10^{-4}\smu$ in the 2mm and 1.3mm bands, respectively. 
We observe an excellent match in the profiles of low-excitation SO lines like $J$=$2_3$--$1_2$ or $J$=$3_4$--$2_3$ and the SiO $J$=2--1 transition (see Fig.~5 and Fig.~A3), showing that SO is tracing the same physical components as SiO.  In the spectra of the 3mm and 2mm transitions (left column in Fig.~A3), the jet signature appears superimposed upon the emission of a broad line component. Proceeding like for the SiO analysis, we find that the jet emission can be fitted by a gaussian peaking at about $+56\kms$ and linewidth (FWHM) of $\approx 9\kms$ (see Table~1). These parameters are similar to those of the SiO jet emission.

As for the transitions in the 1.3mm band (right column in Fig.~A3), a large diversity of profiles is observed, which makes it difficult to disentangle with confidence the contribution of the jet from the broad line component. The situation is more complicated than for SiO. Again, it is quite likely that the beam filling factor of the SO emission is very small and is very sensitive to calibration and pointing errors. We do not take into account these transitions. High-angular resolution with NOEMA are required to clarify this issue.

\subsection{Other species}
Five transitions of H$_2$CO (1 para- and 4 ortho-) were detected at 2mm and 1.3mm. As can be seen in Fig.~A2, the line profiles look all very similar. This is probably because they arise from transitions with similar Einstein coefficient $A_{\rm ij}$ and upper energy level $E_{\rm up}$. Interestingly, the line profiles also look very different from those of SiO and SO, as can be seen in Fig.~1 for instance. The broad line component detected in the former tracers emits very weakly in the H$_2$CO lines, permitting disentanglement of the northern (redshifted) jet emission.

The CS $J$=2--1 and $J$=3--2 lines were detected in the jet (Fig.~A2).  Like for H$_2$CO, the northern jet emission is easily disentangled from the broad line component.

The northern jet signature could be identified in the ground state transition of HCN and HCO$^+$ as a weak component superimposed upon the broad line emission. Unlike the other tracers, the HCN intensity distribution decreases continuously with increasing velocity (Fig.~A4).

\begin{table*}
\caption{Observational and spectroscopic properties of the molecular transitions detected in the northern jet. The observational properties are obtained from a Gaussian fitting to the line profiles. Intensities are expressed in units of antenna temperature corrected for atmospheric absorption ($T_A^{*}$). }
\label{table1}
\begin{tabular}{llrccccccrrrrr}
\hline
      &      &  $\nu$        & $E_{\rm up}$& HPBW      &$\eta_{\rm MB}$ &  W        & $V_0$   & $\Delta v$ & $T_A^*$ & ff\\
      &      & (MHz)         & (K)         &($\arcsec$)&                &($\K\kms$) &($\kms$) &($\kms$) & (mK) & \\ \hline
      &      &               &             &           &($+0\arcsec$,$+12\arcsec$)&         &         &         &      & \\  \hline
CO    & 1--0           &115271.202    & 5.5         & 21.7  & 0.83 & 2.87(.26) & 57.7(.4)& 9.4(1.0)& 290  & 0.11 \\
         & 2--1$^{1}$ &230538.000&16.5         & 21.7  & 0.64 & 10.49(.10)& 56.0(.1)&11.5(.1) & 890  & 0.11 \\
         & 3--2$^{2}$ &345795.990&33.2         & 14.3  & 0.67 & 37.07(.78)& 56.0(.1)&10.4(.3) & $3.35\times 10^3$ & 0.17 \\
         & 4--3$^{2}$ &461040.768&55.3         & 10.7  & 0.45 & 42.77(1.70)&56.5(.2)& 9.4(.4) & $4.4\times 10^3$ & 0.21\\ \hline
        &      &               &             &         &  ($+0\arcsec$,$+0\arcsec$)&  &           &          &        &  \\ \hline
CO    & 1--0 & 115271.202 & 5.5         & 21.7  & 0.83 & 2.87(.26) & 57.7(.4)& 9.4(1.0)& 205  & 0.043 \\
         &          &             &       &      &           &         &         &      &          \\
SiO   & 2--1     & 86846.986& 6.3          & 28.6  & 0.85 & 0.33(.01) & 56.5(.2) & 9.4(.4) & 35(2.7) & 0.043 \\
      & 3--2     &130268.687&12.5            & 18.9  & 0.81  & 0.32(.01) & 55.9(.2) & 9.4(.4) & 33(4.2) & 0.043   \\
      & 5--4     &217104.920&31.3            & 11.3  & 0.65 &   -        &  -       &    -     &  < 3.9 (*)& - \\
SO    & $2_2$ -- $1_1$ & 86093.957 &19.3  & 28.6  & 0.85 & 0.102(.010)& 55.5(.5)& 10.9(1.4) & 8.7(1.6) & 0.043\\
      & $2_3$ -- $1_2$ & 99299.886 & 9.2  & 24.8  & 0.83 & 0.450(.008)& 56.6(.1)& 8.6(.2)   & 49.2(1.6)& 0.043 \\
      & $3_2$ -- $2_1$ &109252.181 &21.1  & 22.5  & 0.82 &0.096(.012) & 56.4(.6)& 9.1(1.5)  & 9.9(2.0) & 0.043\\
      & $3_3$ -- $2_2$ &129138.901 &25.5  & 19.1  & 0.81 &0.246(.015) & 57.1(.2)& 8.2(.6)   & 28.2(3.3)& 0.043\\
      & $3_4$ -- $2_3$ &138178.654 &15.9  & 17.8  & 0.80 &0.772(.010) & 56.5(.1)& 8.2(.2)   & 88.2(2.0)& 0.043\\
      & $4_3$ -- $3_2$ &158971.811 &28.7  & 15.5  & 0.76 &0.364(.010) & 56.7(.1)& 7.9(.3)   & 43.1(2.3)& 0.043\\
      & $4_4$ -- $3_3$ &172181.403 &33.8  & 14.3  & 0.74 &0.346(.020) & 53.6(.5)& 8.8(.7)   & 37.1(4.0)& 0.042 \\
      &                &           &      &       &      &      &      &        &        &  \\
CS    &  2--1          & 97.980953 & 7.1  & 25.1  & 0.84 & 0.193(.009) &58.0(.2)&9.6(.5) & 18.6(1.6) & 0.043 \\
      &  3--2          &146.969026 & 14.1 & 16.7  & 0.78 & 0.223(.008) &57.1(.2)&9.4(.4) & 22.3(2.6) & 0.043 \\
      &  5--4          &244935.555 & 35.4 & 10.0  & 0.62 & 0.076(.027) &55.1(2) &5.8(3)  & 12.3(2.1) & 0.031 \\
      &                &           &      &       &      &      &      &        &        &  \\
H$_2$CO&$2_{1,2}$--$1_{1,1}$ &140839.516& 6.8 &17.5& 0.80& 0.371(.008) &57.2(.1)&7.7(.2) & 45.1(2.6) & 0.043 \\
       &$2_{0,2}$--$1_{0,1}$ &145602.951&10.5 &16.9& 0.78& 0.218(.006) &57.5(.1)&7.9(.3) & 25.9(2.1) & 0.043\\
       &$2_{1,1}$--$1_{1,0}$ &150498.335& 7.5 &16.3& 0.76& 0.271(.022) &57.2(.3)&7.3(.7) & 34.8(1.8) & 0.043 \\
       &$3_{1,3}$--$2_{1,2}$ &211211.449&16.9 &11.6& 0.66& 0.202(.030) &56.3(.8)&10.9(1.9)&17.4(3.1) & 0.039 \\
       &$3_{1,2}$--$2_{1,1}$ &225697.772&18.3 &10.9& 0.65& 0.155(.027) &55.6(.4)&8.4(1.6) &17.3(3.8) & 0.036 \\
       &                     &          &     &    &     &             &         &        &      \\
HCO${^+}$ & 1--0             & 89.188526& 4.3 &27.6& 0.85& 0.197(.015) &56.2(.3)&7.9(.7) & 23.5(1.5) & 0.043 \\
          &                  &          &     &       &      &        &        &         \\
HCN       & 1--0             & 88.631602&4.3  &27.7& 0.85& 0.090(.013) &57.9(.5)&7.1(1.0)& 12.1(1.3) & 0.043  \\
          &                  &          &     &       &      &        &        &      \\
\hline\\
\end{tabular}
\flushleft
$^{1}${Convolved at the resolution of the CO $J$=1--0 line.}\\
$^{2}${Observations with the JCMT.}\\
\end{table*}

\begin{table*}
\caption{Velocity-integrated flux  and spectroscopic properties of the molecular transitions detected at 3mm and 2mm in the southern jet. Line fluxes W are integrated between $-150$ and $-100\kms$. Intensities are expressed in units of antenna temperature corrected for atmospheric absorption ($T_A^{*}$). }
\label{table1}
\begin{tabular}{llrccccc}
\hline
      &      &  $\nu$        & $E_{\rm up}$& HPBW      &$\eta_{\rm MB}$ &  W & ff\\
      &      & (MHz)         & (K)         &($\arcsec$)&                &($\K\kms$)          & \\ \hline
CO    & 1--0 & 115271.202    & 5.5         & 21.7     & 0.83 & 0.67 &  0.043 \\
      &      &               &             &          &      &      &         \\
SiO   & 2--1 & 86846.986     & 6.3         & 28.6     & 0.85 & 0.10 & 0.043 \\
      & 3--2 &130268.687     &12.5         & 18.9     & 0.81 & 0.18 & 0.043 \\
      &      &               &             &          &      &      &     \\
SO    &$2_2$ -- $1_1$& 86093.957&19.3      & 28.6     & 0.85 & 0.03 & 0.043\\
      &$2_3$ -- $1_2$& 99299.886& 9.2      & 24.8     & 0.83 & 0.04 & 0.043 \\
      &$3_2$ -- $2_1$&109252.181&21.1      & 22.5     & 0.82 &  -   & 0.043\\
      &$3_3$ -- $2_2$&129138.901&25.5      & 19.1     & 0.81 &  -   & 0.043\\
      &$3_4$ -- $2_3$&138178.654&15.9      & 17.8     & 0.80 & 0.05 &  0.043\\
      &$4_3$ -- $3_2$&158971.811&28.7      & 15.5     & 0.76 & 0.04 & 0.043\\
      &$4_4$ -- $3_3$&172181.403&33.8      & 14.3     & 0.74 & 0.04 & 0.042 \\
      &              &          &          &          &      &      &     \\
CS    &  2--1        & 97.980953& 7.1      & 25.1     & 0.84 &  -   & 0.043 \\
      &  3--2        &146.969026& 14.1     & 16.7     & 0.78 &  -   & 0.043 \\
      &              &          &          &          &      &      &      \\
H$_2$CO&$2_{1,2}$--$1_{1,1}$&140839.516&6.8&17.5      & 0.80 &  -   & 0.043 \\
       &$2_{0,2}$--$1_{0,1}$&145602.951&10.5&16.9     & 0.78 &  -   & 0.043\\
       &$2_{1,1}$--$1_{1,0}$&150498.335&7.5&16.3      & 0.76 &  -   & 0.043 \\
       &                    &          &   &          &      &      &           \\
HCO${^+}$ & 1--0            & 89.188526&4.3&27.6      & 0.85 & 0.39 & 0.043 \\
          &                 &          &   &          &      &      &            \\
HCN       & 1--0            & 88.631602&4.3&27.7      & 0.85 & 0.02 & 0.043  \\
          &                 &          &   &          &      &      &              \\
\hline\\
\end{tabular}\\
\end{table*}

\begin{table*}
\caption{Velocity-integrated emission of the jet tracers in the northern and southern low-velocity outflow cavity lobes. The fluxes are integrated in the velocity intervals [$-20$;$-14$]$\kms$ and [$-8$;$-2$]$\kms$, associated with the southern and northern outflow lobes, respectively. Intensities are expressed in units of antenna temperature corrected for atmospheric absorption ($T_A^{*}$). }
\label{table1}
\begin{tabular}{llrcccccc}
\hline
      &      &  $\nu$    & $E_{\rm up}$& HPBW  &$\eta_{\rm MB}$ &  W([$-20$;$-14$]) & W([$-8$;$-2$]) & ff\\
      &      & (MHz)     & (K)         &($\arcsec$)&                &($\K\kms$) &($\K\kms$)& \\ \hline

CO    & 1--0 & 115271.202 & 5.5        & 21.7  & 0.83 & 11.9 & 9.8  & 0.20 \\
      & 2--1 &230538.000  &16.5        & 21.7  & 0.64 & 23.2 & 20.7 & 0.20 \\
      & 3--2 &345795.990  &33.2        &14.3   & 0.67 & 27.2 & 25.0 & 0.24\\
      &      &            &            &       &      &      &      &     \\
SiO   & 2--1 & 86846.986  & 6.3        & 28.6  & 0.85 & 0.48 & 0.44 & 0.17 \\
      & 3--2 &130268.687  &12.5        & 18.9  & 0.81 & 0.52 & 0.75 & 0.21   \\
      & 5--4 &217104.920  &31.3        & 11.3  & 0.65 & 0.38 & 0.71 & 0.26  \\
      & 6--5 &260518.018  &43.8        & 9.4   & 0.58 & 0.31 & 0.62 & 0.28  \\
      & 7--6 &303926.814  &58.4        & 8.1   & 0.52 & 0.21 & 0.87 & 0.29 \\
      & 8--7 &347330.592  &75.0        & 7.1   & 0.43 & 0.18 & 0.41 & 0.31 \\
      &      &            &            &       &      &      &      &          \\
SO    & $2_2$ -- $1_1$ & 86093.957&19.3& 28.6  & 0.85 & 0.11 & 0.15 & 0.17 \\
      & $2_3$ -- $1_2$ & 99299.886& 9.2& 24.8  & 0.83 & 0.78 & 0.69 & 0.18  \\
      & $3_2$ -- $2_1$ &109252.181&21.1& 22.5  & 0.82 & 0.19 & 0.15 & 0.19 \\
      & $3_3$ -- $2_2$ &129138.901&25.5& 19.1  & 0.81 & 0.26 & 0.34 & 0.21 \\
      & $3_4$ -- $2_3$ &138178.654&15.9& 17.8  & 0.80 & 0.89 & 1.20 & 0.22 \\
      & $4_3$ -- $3_2$ &158971.811 &28.7& 15.5 & 0.76 & 0.32 & 0.52 & 0.23 \\
      & $4_4$ -- $3_3$ &172181.403 &33.8& 14.3 & 0.74 & 0.31 & 0.56 & 0.24 \\
      & $5_4$ -- $4_3$ &206.176005 &38.6& 11.9 & 0.67 & 0.45 & 0.67 & 0.26 \\
      & $5_5$ -- $4_4$ &215.220650 &44.1& 11.4 & 0.66 & 0.27 & 0.86 & 0.26 \\
      & $5_6$ -- $4_5$ &219.949388 &35.0& 11.2 & 0.66 & 0.78 & 1.25 & 0.27 \\
      & $6_5$ -- $5_4$ &251.825759 &50.7& 9.8  & 0.60 & 0.24 & 0.89 & 0.28 \\
      & $6_6$ -- $5_5$ &258.255826 &56.5& 9.5  & 0.58 & 0.20 & 0.72 & 0.29  \\
      & $6_7$ -- $5_6$ &261.843705 &47.6& 9.4  & 0.58 & 0.71 & 1.16 & 0.29 \\
      &                &           &     &      &     &      &      &        \\
CS    &  2--1          & 97.980953 & 7.1 & 25.1 & 0.84& 1.50 & 0.96 & 0.14  \\
      &  3--2          &146.969026 & 14.1 & 16.7& 0.78& 1.59 & 1.58 & 0.18  \\
      &  5--4          &244935.555 & 35.4 & 10.0& 0.62& 0.85 & 1.51 & 0.22  \\
      &  6--5          &293912.089 & 49.4 & 8.4 & 0.52& 0.53 & 2.22 & 0.24  \\
      &  7--6          &342882.854 & 65.8 & 7.2 & 0.43& 0.41 & 0.96 & 0.25   \\
      &                &           &      &     &     &      &      &        \\
H$_2$CO&$2_{1,2}$--$1_{1,1}$&140839.516&6.8&17.5& 0.80& 1.89 & 1.88 & 0.17 \\
       &$2_{0,2}$--$1_{0,1}$&145602.951&10.5&16.9&0.78& 1.47 & 1.13 & 0.18\\
       &$2_{1,1}$--$1_{1,0}$&150498.335& 7.5&16.3&0.76& 1.77 & 1.51 & 0.18 \\
       &$3_{1,3}$--$2_{1,2}$&211211.449&16.9&11.6&0.66& 1.93 & 1.74 & 0.21 \\
       &$3_{1,2}$--$2_{1,1}$&225697.772&18.3&10.9&0.65& 1.33 & 1.80 & 0.22 \\
       &                &          &      &     &     &      &      &        \\
HCO${^+}$ & 1--0        & 89.188526& 4.3  & 27.6& 0.85& 1.88 & 1.13 & 0.17\\
          &             &          &      &     &     &      &      &         \\
HCN       & 1--0        & 88.631602& 4.3  &27.7 & 0.85& 4.37 & 3.65 & 0.17 \\
          &             &          &      &     &     &      &      &      \\
\hline\\
\end{tabular}\\
\end{table*}

\section{Discussion}
\begin{figure}
\includegraphics[width=0.9\columnwidth]{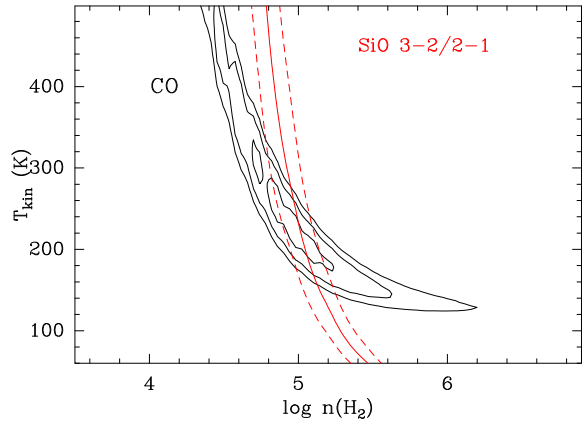}
\caption[]{$\chi^2$ distribution of LVG slab models for the CO emission in the plane ($n(\htwo)$,$T$). Black contour levels of 0.25, 0.50, 0.75 are drawn. The solid red curve draws the observed SiO 3--2/2--1 line intensity ratio as a function of $n(\htwo)$ and $T$. An interval of 20\% around the ratio is delineated by the dashed curves.}
\end{figure}

\subsection{Physical conditions and molecular abundances}
\begin{table}
\centering
\caption{Collisional coefficients adopted for LVG calculations with MADEX}
\label{table1}
\begin{tabular}{lr}
\hline
Species   &   Collisional coefficients  \\ \hline
CO        &   CO--\htwo\ Yang et al. (2010)\\
SiO       &   SiO--p-\htwo\ Dayou \& Balan\c{c}a (2006) \\
SO        &   SO--He Lique et al. (2006) \\
H$_2$CO   &   H$_2$CO--H2 from Green (1991) \\
CS        &   CS--He Lique et al. (2007)   \\
HCN       &   HCN--He Dumouchel et al. (2010)  \\
HCO$^{+}$ &   HCO$^+$--\htwo Flower et al. (1999) \\
\hline
\end{tabular}
\end{table}

We have constrained the  physical conditions and the abundance of the detected molecular species  from the modelling of the line fluxes using the radiative transfer code MADEX in the large velocity gradient (LVG) approximation (Cernicharo 2012). The adopted collisional coefficients are summarised in Table~4. For each molecular species, we have built a grid of models with density between $10^4\cmmt$ and $10^8\cmmt$ and temperature between $30\K$ and $800\K$ to determine the region of minimum $\chi^2$ as a function of  density and temperature.

\begin{table}
\centering
\caption{Physical conditions  determined in the northern and southern lobes for the high-velocity jet and the low-velocity outflow from a LVG analysis of the CO emission with the MADEX radiative transfer code. }
\label{table1}
\begin{tabular}{@{\extracolsep{4pt}}lrrrr@{}}
\hline
                                               & \multicolumn{2}{c}{Jet}                 &  \multicolumn{2}{c}{Outflow} \\
                                               \cline{2-3} \cline{4-5}
                                               &  North                   &   South     & North        & South  \\ \hline
Vel. range ($\kms$) 	& +50;+80	  & -150;-100 &  -8;-2	& -20;-14	\\ \hline
$T_{kin} (\K)$                        &     180-300          &  80-100    &    50          &     50  \\
$n(\htwo) (10^5\cmmt)$      &  0.6--2.0              &  0.5--1.0  &   1--4        &   1--7 \\
$N$(CO)     ($10^{17}\cmmd)$&    2.7                   &  0.9            &  1.2           &   1.4 \\
\hline
\end{tabular} \\
\flushleft
\end{table}

\subsubsection{The jet}

\begin{table}
\centering
\caption{Molecular gas column densities determined in the northern and southern lobes for the high-velocity jet and the low-velocity outflow from a LVG analysis with MADEX.}
\label{table1}
\begin{tabular}{@{\extracolsep{4pt}}lrrrr@{}}
\hline
                      & \multicolumn{2}{c}{Jet}       &  \multicolumn{2}{c}{Outflow} \\ 
                      \cline{2-3}\cline{4-5}
                      &  North    & South            & North       & South  \\ \hline
Species               & $N$($\cmmd$) & $N$($\cmmd$)    & $N$($\cmmd$)       & $N$($\cmmd$) \\ \hline
CO                    &  2.7(17)  & 0.9(17)        &    1.2(17)      & 1.4(17)    \\
SiO                   &  1.7(13)  & 1(13)        &     6(12)       & 5.5(12)  \\
SO                    &  1.0(14)  & 1(13)        &  8.5(13)        &  5.5(13)  \\
H$_2$CO               &  1.2(14)  & $< 2(12)$    &   7.0(13)       &  8.1(13) \\
CS                    &  3.3(13)  & $< 3(12)$    &   3.7(13)       &  5.5(13)\\
HCN                   &  3.3(12)  & $< 5(11)$    &  (4.0-7.5)(13)  &  1.0(14) \\
HCO$^{+}$             &  4.7(12)  & 1(13)        &  (5.5-8.0)(12)  &  1.7(13) \\
\hline
\end{tabular}
\end{table}

\begin{table}
\centering
\caption{Molecular abundances with respect to $\htwo$ determined in the northern and southern lobes for the high-velocity jet and the low-velocity outflow, assuming a standard abundance ratio [CO]/[$\htwo$]= $10^{-4}$.}
\label{table1}
\begin{tabular}{@{\extracolsep{4pt}}lrrrr@{}}
\hline
                      & \multicolumn{2}{c}{Jet}       &  \multicolumn{2}{c}{Outflow} \\ 
                       \cline{2-3}\cline{4-5}                       
                      &  North    & South            & North         & South  \\ \hline
Species               & $X$ & $X$    & $X$       & $X$ \\
\hline
SiO                   &  0.6(-8)  & 1(-8)          &   0.5(-8)       &  0.4(-8)  \\
SO                    &  3.7(-8)  & 1(-8)          &   7.1(-8)       &  3.9(-8)  \\
H$_2$CO               &  4.4(-8)  & $< 0.2(-8)$    &   5.8(-8)       &  5.8(-8) \\
CS                    &  1.2(-8)  & $< 0.3(-8)$    &   3.1(-8)       &  3.9(-8)\\
HCN                   &  0.12(-8)  & $< 0.05(-8)$  & (3.3--6.3)(-8)  &  7.1(-8) \\
HCO$^{+}$             &  0.17(-8)  & 1(-8)         & (0.46--0.67)(-8)&  1.2(-8) \\
\hline
\end{tabular}
\end{table}

In the northern jet,  the parameters of the molecular transitions (flux, linewidth)  were obtained from a simple Gaussian fit to the line profiles, as explained in the previous Section. The results of those fits are summarized in Table~1.
In order to derive the intrinsic brightness temperatures of the molecular transitions detected in the jet, the main-beam brightness temperature obtained from our simple Gaussian fit analysis was corrected for the coupling between the jet and the telescope main-beam.  The latter was estimated following the approach described in Lefloch et al. (2015) in their analysis of the southern lobe of the CO jet.  In a first step, the solid angle of the jet is estimated from the intersection between the half-power contour of the CO distribution mapped with the PdBI and the telescope main-beam. The latter is modeled by a disk centered at the position of the protostar (0$\arcsec$,0$\arcsec$), of diameter equal to the telescope HPBW.
In a second step, the corresponding beam filling factor is obtained from the ratio of the source solid angle to the telescope main beam. In our LVG calculations, we have adopted a linewidth $\Delta v$=$10\kms$ for CO, CS and SiO, and $\Delta v$=$8\kms$ for H$_2$CO and SO, in agreement with the results of  our Gaussian fits to the line profiles (Table~1). 
Our PdBI CO map shows that the northern jet emission is detected from a few arcsec away from the protostar (Fig.~1). The pointing uncertainties for the IRAM 30m telescope, which are typically $2\arcsec$--$3\arcsec$, cast some uncertainties on the actual beam filling factor for the observations in the 1.3mm band, when the telescope beam (HPBW) is $\approx 10\arcsec$ and the covered jet length becomes smaller than the pointing uncertainties. For this reason, we have considered only the millimeter transitions detected in the 3mm and 2mm bands.

In the southern jet, molecular line profiles, like e.g.  SiO $J$=2--1, deviate strongly from gaussianity (see also Fig.~3). For this reason, we have taken into account only the  flux integrated in the velocity range between $-150$ and $-100\kms$. 
Like for the northern jet, we have considered only the molecular transitions detected in the 3mm and 2mm bands, and for each species we have determined the molecular gas column density adopting the kinetic temperature  derived from the CO multi-transition analysis of Lefloch et al. (2015). 
The velocity-integrated fluxes are reported in Table~2. 

Our best fit model of the CO flux in the transitions $J$=1--0 to 4--3 yields a temperature $T_{kin}\approx 220\K$, a gas column density $N$(CO)= $2.7\times 10^{17}\cmmd$, and a gas density $n(\htwo)$ $\approx 1.0\times 10^5\cmmt$. We report in Fig.~6 our $\chi^2$ analysis of the solutions in the plane ($n$,$T$).  Degeneracy is observed and possible solutions are found in the range  $T_{\rm kin}$= 180--$300\K$ and $n(\htwo)$=(0.6--2.0)$\times 10^{5}\cmmt$. Those observations were carried out towards the offset position ($0\arcsec$,$+12\arcsec$) but the beam size is large enough to encompass material close to the protostar (Figs.~1--2).  These physical conditions  should represent a good approximation to those at the origin of the jet. Indeed, our radiative transfer code predicts a CO $J$=1--0 line intensity in reasonable agreement with the observations, when taking into account the jet-beam coupling. In what follows, we will adopt these values for the jet physical parameters in the direction of the protostar. They are summarized in Table~5. 

Analysis of the SO emission brings a very good constraint on the \htwo\ gas density, with $n(\htwo)$= (1--2)$\times 10^5\cmmt$.
Analysis of the CS, SiO and H$_2$CO lines are consistent with these results. This is illustrated in Fig.~6, in which we have reported the value of the SiO 3--2/2--1 line intensity ratio in the parameter space ($n$,$T$). This ratio is a good probe of $n(\htwo)$, whereas it is barely sensitive to the kinetic temperature. In the case of HCN and HCO$^+$, for which only one transition was detected, we ran our LVG code to determine the gas column density, adopting a kinetic temperature $T$=$220\K$ and a gas density $n(\htwo)$=$10^5\cmmt$.

The derived column densities are summarized in Table~6.  The molecular abundances with respect to $\htwo$ were obtained from the ratio of the derived column densities  to that of  CO, assuming a standard abundance ratio [CO]/[$\htwo$]=$10^{-4}$. They are summarized in Table~7.

In  the southern jet, the gas density $n(\htwo)$ derived from the SiO lines is of the order of (2--4)$\times 10^5\cmmt$.
An upper limit was estimated for the species missed in the southern jet: HCN, CS and H$_2$CO.  The results are summarized in Tables~5--7. Despite the difference in column densities between the northern and southern jet, it appears that all the molecular species display similar abundances, of the order of $10^{-8}$ with respect to $\htwo$. This point is further addressed in the next section.

\subsubsection{The low-velocity outflow}

We have estimated the molecular abundances of the various jet tracers in both low-velocity outflow lobes following the same methodology as for the jet. 
In order to avoid contamination of the outflow emission by the hot corino whose line profiles have a typical linewidth of $5$ -- $6\kms$ (see Ospina-Zamudio et al. 2018) and peak at the ambient cloud velocity $v_{lsr}$= $-10.9\kms$, the outflow molecular line fluxes were measured in the velocity intervals [-20;-14]$\kms$ and [-8;-2]$\kms$ for the southern and northern lobes, respectively. The velocity-integrated fluxes measured in both outflow lobes are summarized in Table~3.

The filling factor was obtained from computing the convolution of the telescope main beam with the brightness distribution of the outflow cavity. We assumed that the latter is uniform and the shape of the outflow cavity close to the protostar can be modelled as  a parabola $z$= $0.44 r^2$ where the height $z$ and radius $r$ are expressed in arcsec units (see Fig.~B1). 

Our LVG analysis of the CO transitions $J$=1--0, 2--1 and 3--2 indicates similar physical conditions in both lobes with a gas kinetic temperature $T_{kin}$ $\simeq 50\K$ and a column density $N$(CO)$\simeq 1.2\times 10^{17}\cmmd$.  Multi-line analysis of SO and CS yield consistent, typical $n(\htwo)$ densities in the range (1--4)$\times 10^5\cmmt$ and (1--8)$\times 10^5\cmmt$ in the northern and southern outflow lobes, respectively. These results are consistent with the physical conditions previously estimated towards the southern outflow lobe by Lefloch et al. (2015).

In the case of the SiO emission, it is not possible to fit the whole spectral line energy distribution (SLED) with one single gas component.  We note that Lefloch et al. (2015) reported evidence for a gradient of excitation conditions in the outflow (see their Table~4). Therefore, we have modelled only the emission of the lower-excitation SiO lines $J$=2--1 and 3--2. We obtained  typical  $n(\htwo)$ densities of a few  $10^5\cmmt$, consistent with the SO and CS analysis.  

In the case of H$_2$CO, we modelled separately the emission of the ortho- para- lines.  All four ortho transitions $2_{1,2}$--$1_{1,1}$, $2_{1,1}$--$1_{1,0}$, $3_{1,3}$--$2_{1,2}$ and $3_{1,2}$--$2_{1,1}$  were fitted simultaneously. We then adopted the same kinetic temperature and \htwo\ density and we determined the column density $N$(p-H$_2$CO) necessary to reproduce the flux of the para transition $2_{0,2}$--$1_{0,1}$. We obtain an ortho to para ratio of 2.5, which is compatible with the canonical value of 3, taking into account the flux uncertainties.

The outflow physical conditions, molecular gas column densities and  abundances are summarized in Tables ~5, 6 and 7, respectively.

\subsection{Chemical differentiation}

How different are  the physical and chemical conditions between the northern and the southern jets ?
The physical conditions estimated in the northern jet indicate a higher temperature (\textit{T}$\simeq$200$\K$) and a higher CO column density ($N$(CO)$\simeq$3$\times 10^{17}\cmmd$) than in the southern lobe (\textit{T}$\simeq$100$\K$ and $N$(CO)$\simeq$1$\times 10^{17}\cmmd$).
The high-velocity CO gas distribution, as observed with the PdBI (Fig.~1) displays the same "knotty" structure in both lobes. 
Lefloch et al. (2015) interpreted these knots as the signature of internal shocks  which act as a local source of gas heating in the jet, and are responsible for the high temperature ($750-1200\K$), dense ($\sim 10^6\cmmt$) CO gas component  
detected  in the southern lobe.  We speculate that similar excitation components are also present in the northern jet.

The jet dimensions (length and transverse size) as determined from the brightness contour at half power are found to be similar in both the northern and southern jets, and can be fitted approximately by an ellipse of $20\arcsec \times 1.7\arcsec$, which implies a typical radius of $0.85\arcsec$ or 620~au for the jet. The radius of the high-velocity CO jet does not display significant variations with distance to the protostar and is $\sim6$ times larger that the values reported towards molecular jets of low-mass protostars like HH212 (90AU, Cabrit et al. 2012) or L1157 (60-125AU, Podio et al. 2016). The CO brightness distribution appears rather loose at low intensity levels, suggesting the presence of a complex  structure at subarcsec scale. The Parallactic Angle is different between the two lobes however, with values of $+64\deg$ and $+69\deg$, respectively. We note that the gas distribution extends to larger radii, up to  about $1.7\arcsec$ ($2.0\arcsec$) in the southern (northern) jet, which is suggestive of {\em entrained gas} along the jet. 

In the jet, a comparison of the molecular abundances in the northern and southern lobe shows that they are similar within a factor of 2. Differences in the relative composition are detected, however: the most abundant species in the northern jet are SO and H$_2$CO, with an abundance of about $4\times 10^{-8}$ relative  to $\htwo$; on the contrary, the most abundant species in the southern jet are SO and SiO, with a relative abundance of $2\times 10^{-8}$. Also,  HCO$^{+}$, appears to be more abundant by a factor of 2 in the southern jet. CS and H$_2$CO, which are both undetected in the southern jet, are  underabundant by a factor of 4 {\em at least} with respect to the northern jet.

In the outflow, our analysis  shows that  the physical and chemical conditions in the northern and southern lobes are also very similar. As a matter of fact, the abundances of the different tracers considered in the present work (i.e. those detected in the jet) vary by less than a factor of 2 between the two outflow  lobes (see Table~7).
Our unbiased spectral line survey of CepE-mm shows that the jet is often not detected in the tracers of the chemically-rich outflow gas, which proves a strong chemical differentiation between the jet and the outflow. As an example, none of the complex organic molecules (COMs) detected in the low-velocity outflow (the {\em extremely broad line component} eBL in Ospina-Zamudio et al. 2018) displays any detectable signature in the protostellar jet. 

It is rather surprising that molecular species such as H$_2$CO and CS, which are detected in both outflow lobes with similar abundances, are detected only in the northern jet, with an abundance similar to that of the outflow gas. Conversely, the  abundances of molecular species detected both in the outflow and the jet are similar within a factor of a few (Table~7). The only exception is HCN, which is found much lower in the jet by one order of magnitude. We must take care in not overinterpreting these quantitative results as we cannot exclude that  the emission of the different tracers  may arises from  different regions of the jet.  The question arises, however, as to whether such a chemical differentiation is  caused by the different physical conditions between the jet and the outflow or by a fundamentally different origin between  the high-velocity material and the low-velocity outflow gas.

\subsection{Origin of the emission}
\subsubsection{Evidence for jet entrained gas}
Analysis of the molecular line emission reveals the presence of several species with abundances of the order of $10^{-8}$. The detection of these various molecular species makes Cep\,E-mm one of the richest molecular jets reported until now, together with L1448 and I04166 (Tafalla et al. 2010).
Several observational facts are intriguing and are difficult to reconcile with an origin of the molecular emission in the protostellar jet. First, the jet diameter derived from the CO emission is about one order of magnitude larger than the values measured in typical low-mass Class 0 jets or predicted by models, which are usually of the order of a few times 10 au. This suggests that CO is not tracing the protostellar jet alone. Second, some of the most abundant species, like H$_2$CO ($4.4\times 10^{-8}$ in the northern jet) or HCO$^{+}$ ($10^{-8}$ in the southern jet) are  usually not detected in young protostellar jets. Last, a strong degree of chemical differentiation is observed between the northern and southern jets, which is difficult to understand if the jet is made of pristine material taken away from the protostellar disk or formed in the jet itself.

Therefore, it is quite likely that the  emission detected in our spectral survey does  {\em not} originate from the protostellar jet itself. We note that both the jet and the low-velocity outflow are  dense, with a density $n(\htwo)\approx $ a few times $10^5\cmmt$ in the jet up to $\sim 10 ^6\cmmt$ in the entrained gas of the outflow. Under such conditions, the formation of a turbulent molecular layer at the interface between the jet and outflow cavity gas is to be expected.

The results of our study are biased by the lack of angular resolution of the IRAM 30m spectral line survey and we cannot exclude that different species are actually tracing different regions of the jet/outflow system. This hypothesis is supported by the differences observed between the line profiles of the molecular tracers (see Fig.~1). Interferometric observations at arcsec scale are required to clarify this point and better understand the chemical differentiation along the jet.

\subsubsection{Shock-driven photochemistry at work?}

Several of the measured molecular abundances are similar in the jet and the outflow and it is not clear whether this similarity is coincidental or it reflects the nature of the molecular-rich high-velocity gas, i.e. entrained gas from the outflow cavity. Even so, this hypothesis does not explain why some species are present in one lobe and not the other, as illustrated by HCO$^{+}$ and H$_2$CO. Conversely, the HCN abundance in the jet is about one order of magnitude less than in the low-velocity outflow. 

We propose that the unusual chemistry observed in/around the CepE-mm could be driven by internal shocks along the jet or in the turbulent layer of outflow gas entrained by the protostellar jet.
As discussed by Lefloch et al. (2015) and in this work, there is strong evidence for the propagation of shocks
and instabilities along the jet. The dynamical age of the jet is short enough, of the order of $1000\yr$, that shocks are expected to display mixed C- and J-type components. We speculate  that these shock components create a diffuse UV radiation field which will also illuminate the surrounding outflow gas.

Viti et al. (2002) have investigated the impact of such a shock-driven UV field on the chemical composition of the parental cloud gas at the interface with the outflow. Their modelling cannot be directly applied to the present situation as the chemical and physical conditions in the outflow gas are different from the quiescent envelope considered by them. On the one hand, the outflow gas is warmer than the envelope, and, on the other hand, dust grains have already been processed and a fraction of their icy mantles eroded through the bowshock responsible for the formation of the outflow. However, this modelling can bring some insight on processes at work in the Cep\,E-mm outflow/jet system.  In their modelling, Viti et al (2002) consider a shock producing a UV radiation field with an intensity of $\chi$= 10 times the Habing radiation field (Habing, 1968). The shock-induced photoionization  drives a rich  photochemistry in the immediate vicinity of the outflow and progresses into the core material. As can be seen in their Figure~2, H$_2$CO and SO are produced with similarly high abundances ($\sim 10^{-7}$) and remain for several $10\yr$ in the gas phase. 
Interestingly, they find that right after CO and H$_2$O, the most abundant species is HCO$^+$ (abundances $\sim$ a few $10^{-9}$), provided that the surrounding gas temperature is not too high ($30$ -- $100\K$).
A higher temperature ($\sim 1000\K$) extends the duration of the chemically-rich phase, resulting in HCO$^{+}$ abundance of $\sim 10^{-8}$.
Interestingly,  the predictions of this model appear to agree qualitatively with our observations. This agreement is very encouraging and supports the scenario of a shock-driven chemistry in a turbulent layer around the high-velocity jet. Interferometric observations at subarcsec scale are required to elucidate the chemical differentiation observed along the jet and the behaviour of the different tracers, so to confirm or rule out the role of shock-driven chemistry in the outflow.

\section{Conclusions}
We have carried out a systematic search for molecular line emission from the Cep\,E-mm high-velocity protostellar jet using the IRAM 30m telescope, complemented with JCMT and PdBI observations. In addition to CO,  we have identified several molecular species in the jet: SO, SiO, H$_2$CO, CS, HCO$^{+}$ and HCN. We find evidence for chemical differentiation between the northern and southern jet lobes. Whereas all these aforementioned species are detected in the northern jet, only CO, SO, SiO and HCO$^{+}$ are detected in the southern jet.

A radiative transfer analysis of the molecular line emission in the Large Velocity Gradient approximation with MADEX has permitted us to constrain the physical and chemical conditions in the jet and the low-velocity outflow. We find the northern jet is made of hot ($180-300\K$), dense ((0.6-2.0)$\times 10^5\cmmt$) gas. Overall, the jet material appears warmer and denser in the northern than in the southern jet, where a kinetic temperature of 80--$100\K$ and density of (0.5--1.0)$\times 10^5\cmmt$ were determined by Lefloch et al. (2015). Also, the jet appears warmer and less dense than the entrained outflowing gas, with $T_{kin}\simeq 50\K$ and $n(\htwo)$= (1--7)$\times 10^5\cmmt$. Interestingly, the physical conditions in the low-velocity gas are very similar in both outflow lobes.

The molecular abundances derived in the jet are of the order of $10^{-8}$, and are found to be similar to those determined in the low-velocity outflow gas. They reveal an unusual chemical composition with respect to that commonly observed in protostellar jets. A strong chemical differentiation is observed between the two jet lobes. In the northern jet, we find that H$_2$CO and SO are the most abundant species, immediately after CO. In the southern jet, the most abundant species are SO, SiO and HCO$^{+}$.

The transverse size of the CO jet was estimated $\approx$ 1200~au from Plateau de Bure observations at $1\arcsec$ angular resolution. This is much larger than the values reported  in protostellar jets and suggests that the jet entrains a turbulent gas envelope in the course of its propagation. 

A qualitative agreement is observed between the jet chemical composition and the predictions of a simple model of a turbulent layer at the ouflow/cloud interface by Viti et al. (2002).  We propose that the detected molecular species are the signatures of the specific photochemistry driven by the UV field generated by shocks in the turbulent envelope. 

Interferometric observations at arcsec scale are required to elucidate the origin of the jet molecular emission and to understand the observed differentiation in the protostellar gas.

\section*{acknowledgements}

We acknowledge funding from the European Research Council (ERC) under the European Union's Horizon 2020 research and innovation programme, for the Project "The Dawn of Organic Chemistry" (DOC), grant agreement No 741002.
Based on observations carried out under project number 033-16 with the IRAM 30m telescope. IRAM is supported by INSU/CNRS (France), MPG (Germany) and IGN (Spain). This work  was supported by a grant from LabeX Osug@2020 (Investissements d'avenir - ANR10LABX56). This work has made use of  MADEX (Cernicharo, 2012).

\appendix
\section{Molecular transitions detected in the protostellar jet of Cep\,E-mm.}

Fig.~A1--A4 present panels of all the detected transitions of SiO, CS, H$_2$CO, SO, HCN and HCO$^+$ in the protostellar jet of Cep\,E-mm.

\begin{figure}
\centering
\includegraphics[width=0.8\columnwidth]{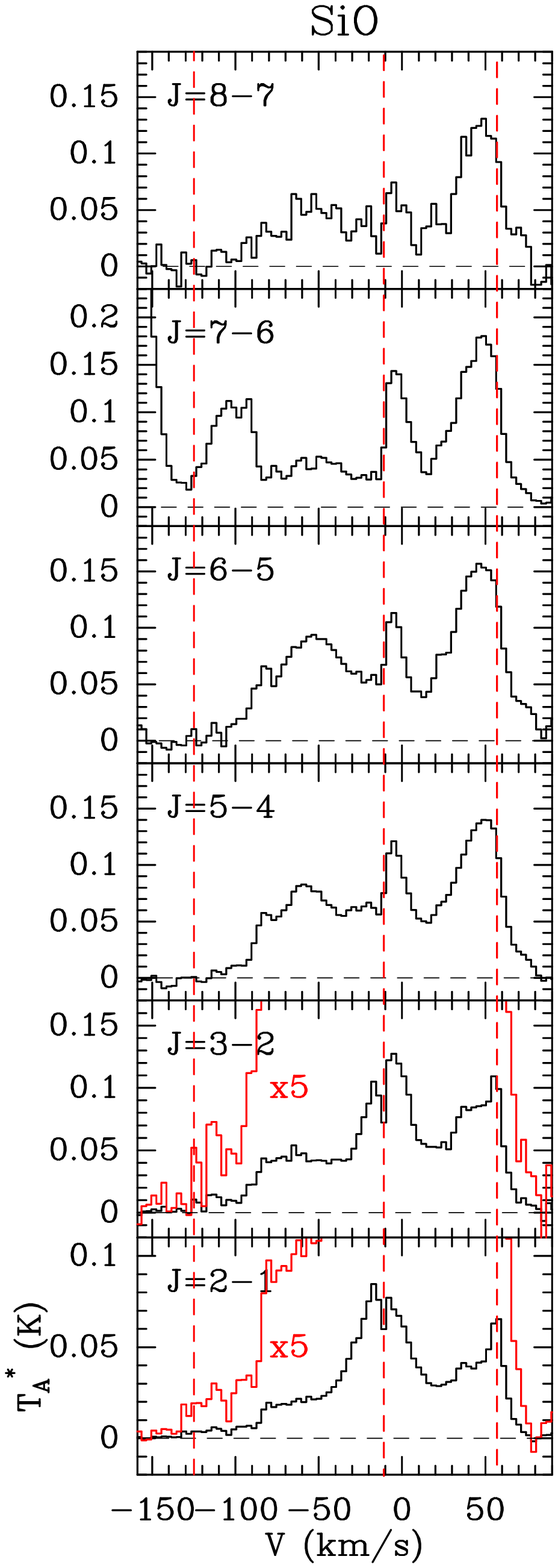}
\caption{Montage of the SiO transitions detected with the IRAM 30m telescope. The jet emission peak velocity in the blue and redshifted lobe at $v_{lsr}$= $+57\kms$ and $-125\kms$, respectively, is marked with a red dashed line. Fluxes are expressed in units of antenna temperature $T_{\rm A}^{*}$, corrected for atmospheric absorption.
}
    \label{panel_line}
\end{figure}

\begin{figure}
\centering
\includegraphics[width=0.49\columnwidth]{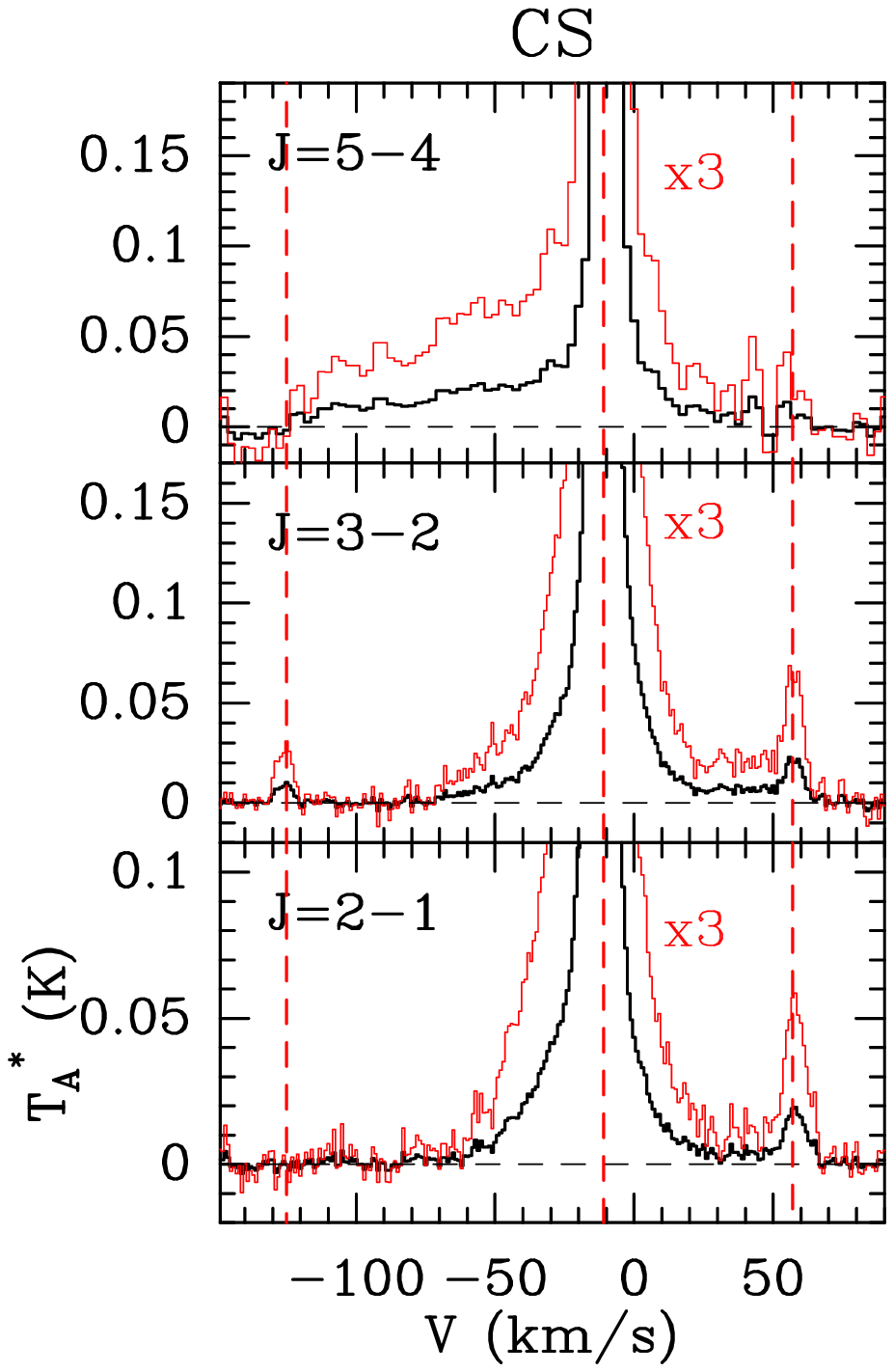}
\includegraphics[width=0.49\columnwidth]{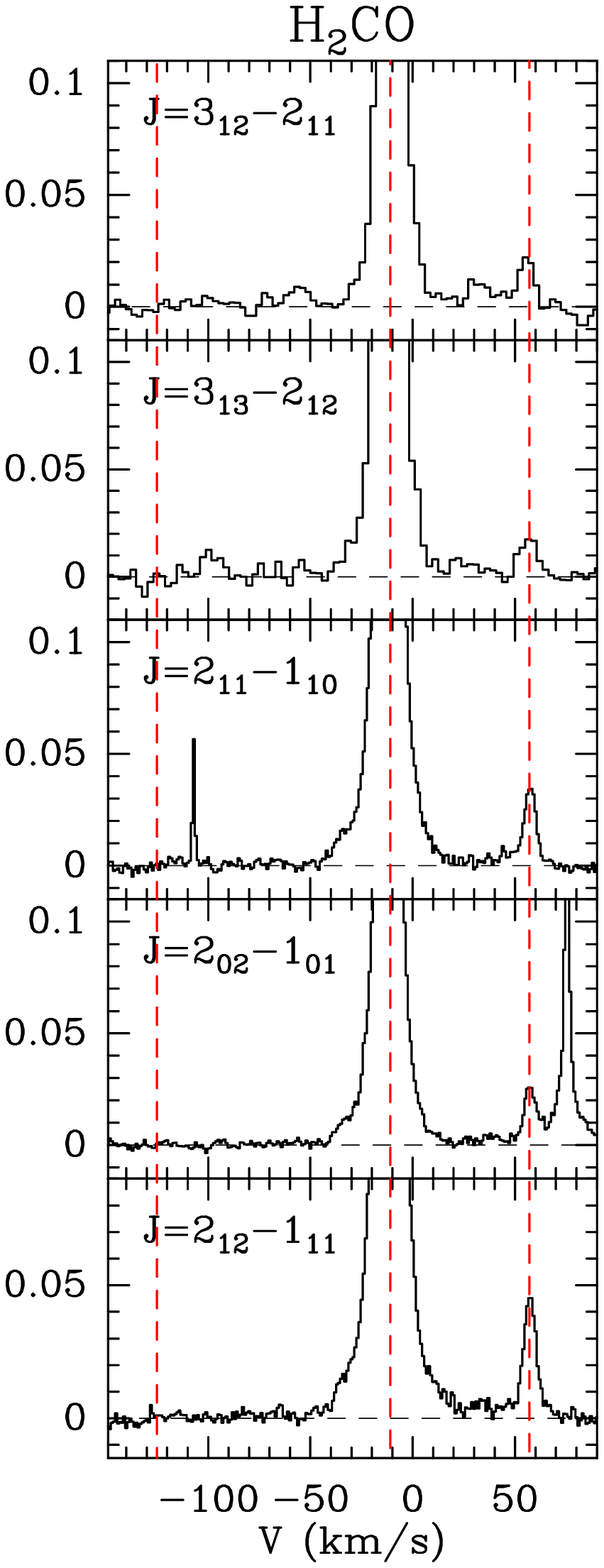}
\caption{{\em(left)}~Montage of the CS transitions (left) and  H$_2$CO transitions (right) detected with the IRAM 30m telescope. The jet emission peak velocity in the blue- and redshifted lobe, at $v_{lsr}$= $-125\kms$ and $+57\kms$, respectively, is marked with a red dashed line. Fluxes are expressed in units of antenna temperature $T_{\rm A}^{*}$, corrected for atmospheric absorption.
}
    \label{panel_cs}
\end{figure}

\begin{figure}
\centering
\includegraphics[width=\columnwidth]{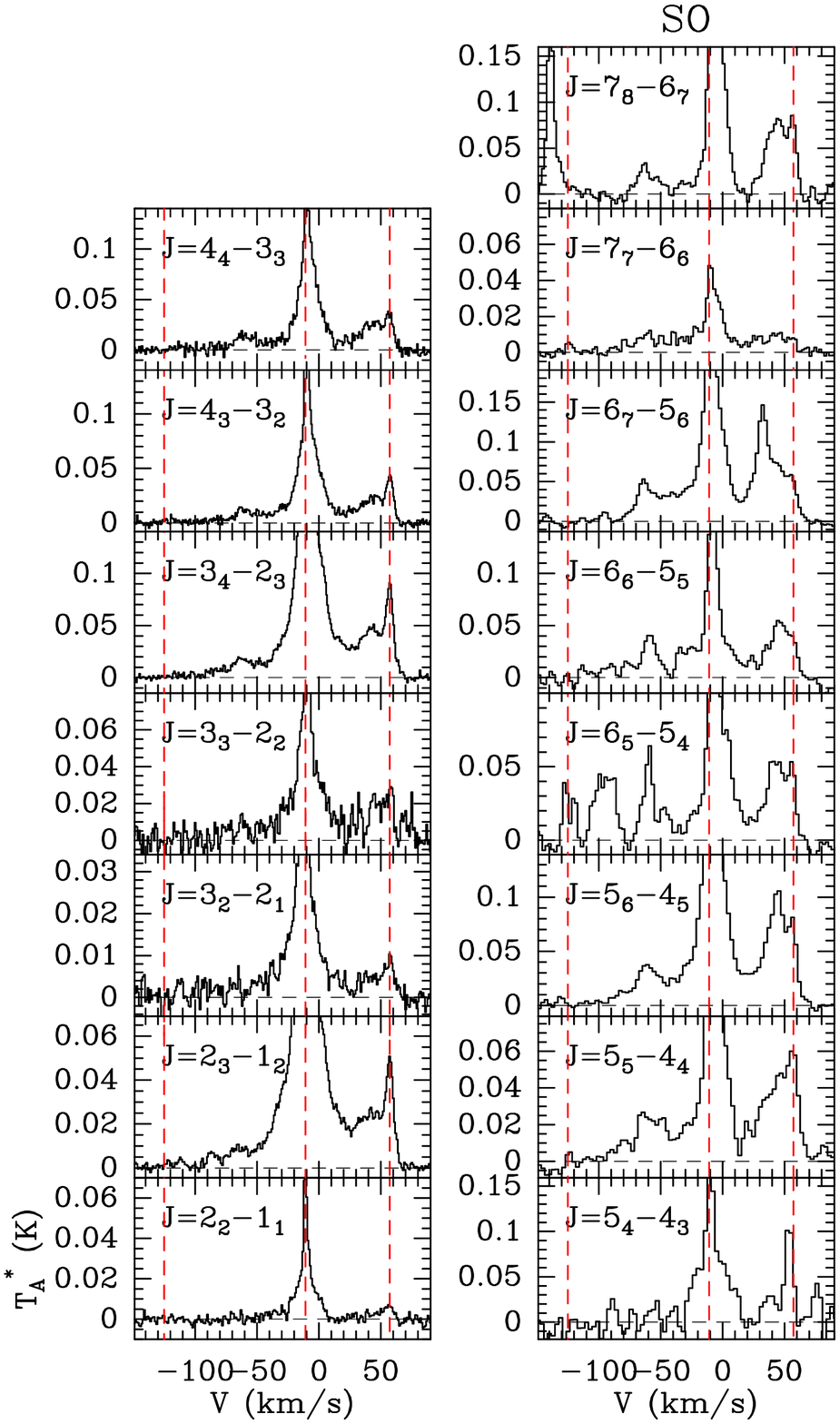}
\caption{Montage of the SO transitions detected with the IRAM 30m telescope. The jet emission peak velocity in the blue- and redshifted lobe, at $v_{lsr}$= $-125\kms$ and $+57\kms$, respectively, is marked with a red dashed line. Fluxes are expressed in units of antenna temperature $T_{\rm A}^{*}$, corrected for atmospheric absorption.
}
\label{panel_so}
\end{figure}

\begin{figure}
\centering
\includegraphics[width=\columnwidth]{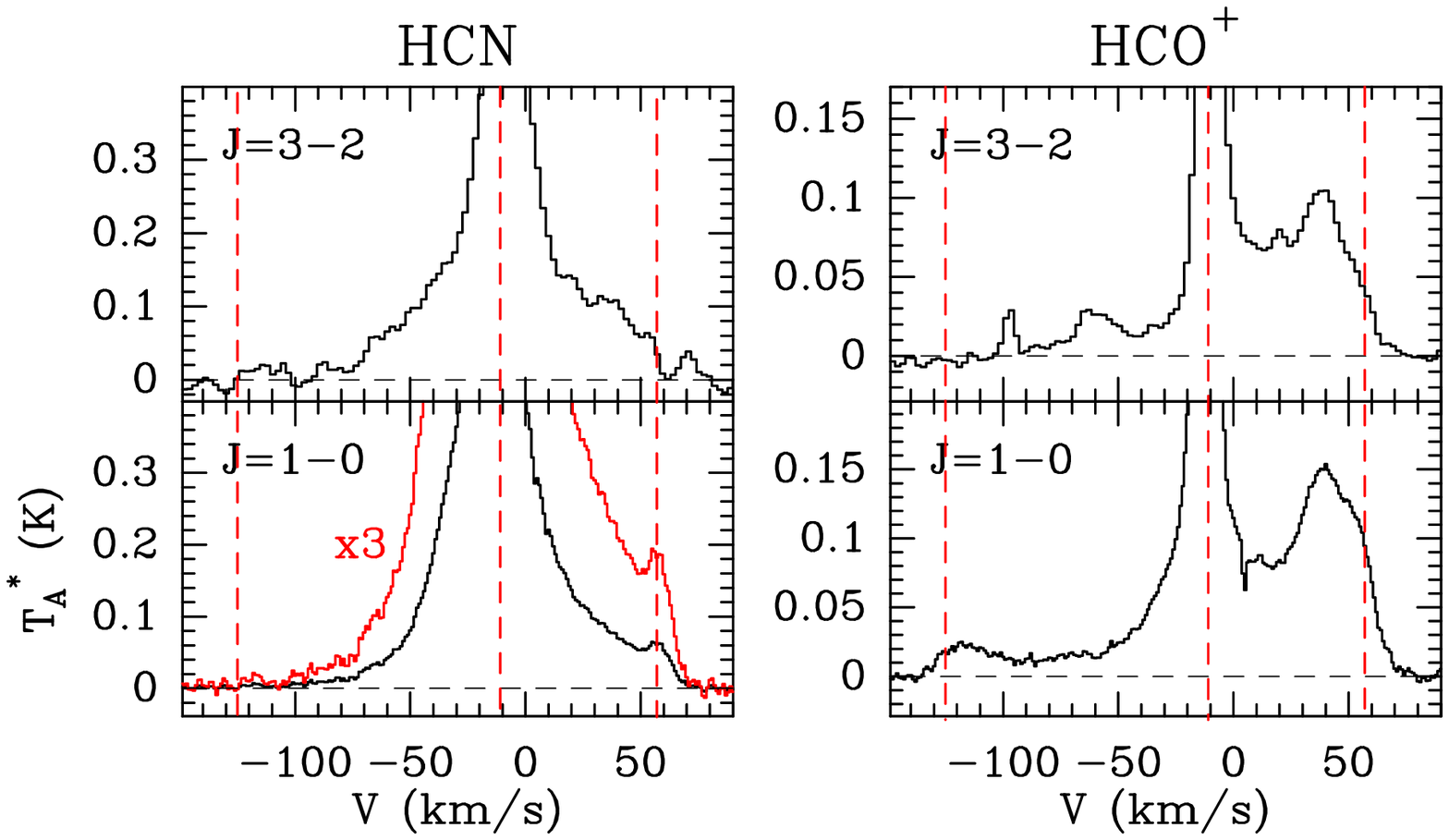}
\caption{ {\em (Left)} Montage of the HCN transitions detected with the IRAM 30m telescope. The jet emission peak velocity in the blue- and redshifted lobe, at $v_{lsr}$= $-125\kms$ and $+57\kms$, respectively, is marked with a red dashed line. {\em (Right)} Montage of the HCO$^+$ transitions detected with the IRAM 30m telescope. The jet emission peak velocity in the blue- and redshifted lobe, at $v_{lsr}$= $-125\kms$ and $+57\kms$, respectively, is marked with a red dashed line. Fluxes are expressed in units of antenna temperature $T_{\rm A}^{*}$, corrected for atmospheric absorption.}
\label{panel_HCOP_HCN}
\end{figure}

\section{Modelling of the outflow cavities}

Modelling of the outflow cavities brightness distribution was performed from the PdBI map in the CO $J$=2--1 line. We assumed that the outflow brightness distribution is uniform and that the shape of the cavities close to the protostar can be modelled as a parabola $z$= $0.44 r^2$ where the height $z$ and radius $r$ are expressed in arcsec units (see Fig.~B1).

\begin{figure}
\includegraphics[width=\columnwidth]{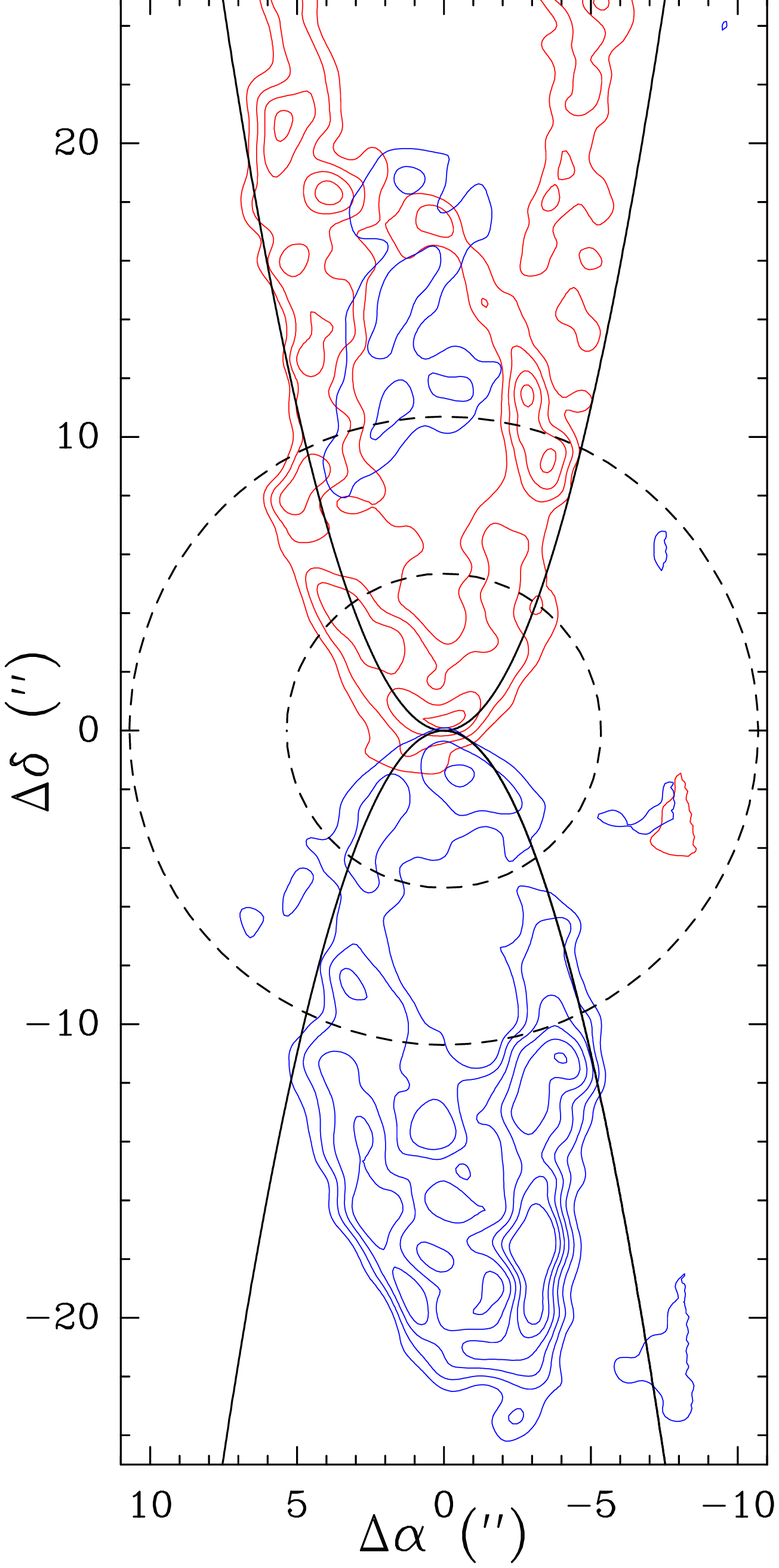}
\caption{CO $J$=2--1 emission in the low velocity outflow cavity as observed with the Plateau de Bure interferometer at $1\arcsec$ resolution. The emission is displayed in a rotated frame so that the main axis of each outflow cavity coincides with the $\delta$-axis. In the northern lobe (red contours) the flux is integrated between  $-8$ and $-2\kms$ and the rotation angle is $12\deg$.  In the southern lobe (blue contours), the flux is integrated between $-20$ and $-14\kms$ and the rotation angle is $20\deg$. 
First contour and contour interval are 1.3 Jy/beam$\kms$ in both lobes. 
The fit to the outflow cavities $\delta$= $\pm 0.44 \alpha^2$ is superimposed in black. The size of the IRAM 30m main beam (HPBW) at the frequency of the CO $J$=1--0 and 2--1 lines drawn by the circles in dashed. 
}
\end{figure}

\end{document}